# Zero-Energy Modes on Superconducting Bismuth Islands Deposited on Fe(Te,Se)

Xiaoyu Chen*, Mingyang Chen*, Wen Duan, Huan Yang, and Hai-Hu Wen†

National Laboratory of Solid State Microstructures and Department of Physics, Center for Superconducting Physics and Materials, Collaborative Innovation Center for Advanced Microstructures, Nanjing University, Nanjing 210093, China

*These authors contributed equally to this work.

† Correspondence and requests for materials should be addressed to hhwen@nju.edu.cn.

**Topological superconductivity is one of the frontier research directions in condensed matter physics. One of the unique elementary excitations in topological superconducting state is the Majorana fermion (mode) which is its own antiparticle and obeys the non-Abelian statistics, and thus useful for constructing the fault-tolerant quantum computing. The evidence for Majorana fermions (mode) in condensed matter state is now quickly accumulated. Here we report the easily achievable zero-energy mode on the tunneling spectra on Bi islands deposited on the Fe(Te,Se) superconducting single crystals. We interpret this result as the consequence of proximity effect induced topological superconductivity on the Bi islands with strong spin-orbital coupling effect. The zero-energy mode is**

**argued to be the signature of the Majorana modes in this size confined system.**

Topological superconductivity is very interesting for several reasons. Firstly, the topological superconducting state should possess by itself the spin-triplet component which is very rare in nature. It has been argued that the superconductivity in $Sr_2RuO_4$ *(1)* and superfluidity in $^3He$ *(2)* may have topological nature because the Cooper pair is a triplet and the gap function behaves as $p_x + ip_y$. Another reason is that an exotic elementary excitation termed as Majorana zero mode (MZM) can arise from the topological superconducting state. The MZM has a characteristic that it is its own antiparticle, such excitations obey non-Abelian statistics, and this is useful for making the fault-tolerant quantum computation *(3-5)*. However, the MZM is also quite rare in ordinary world. It has been argued that in three dimensional (3D) topological superconductors, one can find the MZM at the center in the vortex cores *(6,7)* or at the terminal of a one dimensional (1D) topological superconductor *(8)*. Theoretically it has been proposed that the topological superconductivity may be effectively induced by the proximity effect between an *s*-wave superconductor and a topological insulator *(9)*, in this case the vortex core has a MZM, due to spin-orbital coupling (SOC) that modifies the Caroli-de Gennes-Matricon spectrum. In the self-doped state of $Cu_xBi_2Se_3$, zero-bias conductance peak (ZBCP) was observed on the surface of the sample by the point contact tunneling measurements *(10)*. Later on the scanning tunneling microscopy measurements reveal a full gap feature, which seems to be at odd with the scenario of Majorana mode in this system *(11)*. By placing a nano-wire in adjacent to an *s*-wave superconductor, and by using a perpendicular magnetic field to induce the Zeeman effect, the Delft group produces an 1D topological superconducting state. ZBCPs were observed at the two terminals when a suitable electric field is applied to the gating device *(12)*. Observation of the MZM were achieved in a more straightforward way by attaching an 1D magnetic chain to the s-wave superconductor *(13,14)*. By depositing $Bi_2Te_3$ on the conventional superconductor 2H-NbSe$_2$, superconductivity was induced and a spin (field vorticity) dependent ZBCP was observed within the vortex core *(15,16)*. Recently, in iron based superconductor Fe(Te,Se), the ZBCP was observed within the vortex cores *(17,18)*. Similar phenomenon was found in another iron based superconductor $(Li_{1-x}Fe_x)OHFeSe$ *(19)*. Nematic superconductivity was also observed on the $Bi_2Te_3$/Fe(Te,Se) heterostructure, and a ZBCP was observed within the elongated vortex core and attributed to the presence of Majorana modes *(20)*. In this paper, we report the evidence of easily achievable zero-energy mode which is the signature of Majorana mode in some Bi islands deposited on the iron based superconductor Fe(Te,Se). The ZBCP is robust and exists almost everywhere on the selected Bi islands. We argue that this is the consequence of topological superconductivity induced by the proximity effect between the superconductor Fe(Te,Se) and Bi islands with the strong SOC effect.

The atomically resolved surface of the Fe(Te,Se) substrate shows a square lattice structure (Fig. 1B) with the white and dark spots representing the Te and Se atoms, respectively (*21-23*). We deposited Bi on the surface of Fe(Te,Se) substrate. The deposited Bi elements form some islands which distribute randomly on the surface of Fe(Te,Se) (Fig. 1A), and they have different sizes, shapes and heights. Some of them show an elongated rectangular structure, while the others are roughly rounded. Most of the islands have the height of about 7 Å (e.g., Bi island #1 in Fig. 1C), while some others have a height of about 10 Å (e.g., Bi island #2 in Fig. 1E) or 14 Å (e.g. the most bright bar like islands in Fig.1A). For the islands with the height of about 7 Å, the Bi atoms on the island surface seem to exhibit a locally distorted orthorhombic structure (Fig. 1D), which is similar to the case of ultra-thin Bi film with a slipped Bi(110) structure *(24)*. For the island with the height of about 10 Å, the surface structure is hexagonal with a lattice constant of about 4.5 Å (Fig. 1F), which is also confirmed by six-fold symmetric Bragg peaks in the Fourier-transformed topography (Fig. 1F, inset). And such surface is likely to be the Bi(111) layer. Previous work *(25)* reveals that epitaxial Bi thin films exhibit a transition of growth manner from Bi(110) to Bi(111). On the islands with round shape and diameter of about 6-8 nanometer, even the surface structures are different (Fig. 1D and Fig. 1F), we all observed the ZBCP.

Tunneling spectra measured at 0.4 K on the Fe(Te,Se) substrate show fully gapped feature with a pair of coherence peaks at about ± 2 meV (Fig. 2, A and B). On the Bi islands, one can still see the gapped feature on the spectra, which means that superconductivity is successfully induced on the Bi islands by the proximity effect from the superconducting Fe(Te,Se) substrates. In addition, there are obvious in-gap state peaks on the spectra measured on the Bi islands with the height of 7 (Fig. 2A) or 10 Å (Fig. 2B). The peak position is at or near zero bias no matter where the spectra were taken on the Bi islands. This feature can be further proved by the differential conductance mappings at zero bias (Fig. 2, C and D). The differential conductance, which is proportional to the local density of states (DOS), is almost zero at zero bias on the Fe(Te,Se) substrate because of the gapped DOS within superconducting gap. In contrast, ZBCPs exist at positions all around the Bi islands. On the Bi islands when the bias voltage is at ± 0.8 mV which is still within the superconducting gap, the magnitude of differential conductance becomes strongly suppressed. As shown in Fig.2, D to F, a contrast of the differential conductance images measured at 0 and ± 0.8 mV on island-#2 can be seen. On can see that the image of the ZBC is quite bright, indicating a strong intensity at zero bias.

The emergence of the ZBCP on the Bi islands can be clearly illustrated by the spatial evolution of the tunneling spectra across two Bi islands with different heights (Fig. 3). One can see that, the spectra measured on the Fe(Te,Se) substrate (Fig. 3, blue curves) show a fully gapped feature, but the shapes and energies of the coherence peaks vary with location, which is due to the intrinsic electronic properties in iron-based superconductor. The ZBCP emerges immediately for the spectra measured just at the edges of the Bi islands (Fig. 3, green curves), and it always exists over the whole Bi island (Fig. 3, red curves). On the Bi island #1 with the height of 7 Å, the in-gap state peak energies are stably fixed at zero bias, which is almost independent of the spatial positions (Fig. 3, B and C) although the topographic image (Fig. 1D) is very inhomogeneous. However, the in-gap state peak energy varies slightly when across the Bi island #2 with the height of 10 Å (Fig. 3, E and F) although the top surface of this island has a well ordered hexagonal structure. We must emphasize that the existence of the ZBCP is not a rare phenomenon among the islands, although at this moment we cannot tell what factor and thus structure intimately determines the presence of the ZBCP.

To clarify the intrinsic nature of the ZBCPs, we measured the temperature and magnetic field dependent tunneling spectra on island #2. With increasing temperature, the intensity of the peak weakens and finally disappears at around 5 K (Fig. 4A); meanwhile the superconducting gap feature is suppressed with increase of temperature and finally disappears at 14 K near the bulk $T_c$ of Fe(Te,Se), which is similar to the situation on the Fe(Te,Se) substrate (Fig. 4B). This indicates that the ZBCPs are more sensitive with temperature than the induced superconductivity from proximity effect. This is similar to the case of ZBCP appearing in the vortex core of Fe(Te,Se) *(17,18)*, which was argued as the presence of MZM *(17)*. When applying a magnetic field up to 8 T, the ZBCPs are clearly suppressed but the peak feature still remains (Fig. 4C). And the shift of the peak energies is very small with increase of magnetic field, which excludes the possibility that the in-gap state peaks are originated from the Yu-Shiba-Rusinov state(YSR) induced by magnetic impurities.

We have observed robust ZBCP on some round-shaped Bi islands with the diameter of about 6-8 nanometers. On some islands, the in-gap state peaks locate precisely at zero bias, see for example the island #1 (Fig.3 B and C). The control experiments on islands #3 and #4 also show the ZBCPs (fig. S2). The temperature dependent evolution of the ZBCPs of island #3 and #4 (fig.S3 A and C) look very similar to that of island #2, indicating that the in-gap state peak near zero bias should have the same origin. This ZBCP appears almost everywhere on the selected islands, which excludes the possibility that it is induced by the YSR states. We attribute this ZBCP to the emergence of MZM and topological superconductivity in the selected Bi islands. The p-orbital electrons of Bi atoms couple strongly with the conduction electrons of Fe(Te,Se) and the proximity effect induces superconductivity. Due to the SOC effect, the superconductivity on the Bi island may have triplet component, making the system to be topological in nature. On the surface of the Bi island with topological superconductivity, there are many reasons to generate the MZM. We must emphasize that, within the framework of the present theories for topological superconductor and Majorana modes, it seems difficult to understand the observation of ZBCP almost everywhere on the selected islands. However, as mentioned above, due to the strong spin-orbital coupling of Bi and the proximity effect induced superconductivity, we are confident that the superconducting state on the Bi islands should be topological in nature. This gives strong confidence that the ZBCP is the signature of Majorana mode. Based on the observed features, we would like to propose three scenarios for explaining our results. (1) The Bi island may exhibit the surface Andreev bound states arising from the excitations of the Majorana modes, as proposed for the system of $Cu_xBi_2Se_3$ *(10,26,27)*. According to this theoretical proposal, two conductance peaks close to zero-bias may appear when the measuring temperature is low enough. At finite temperatures, the two peaks may merge into a broad peak near zero-bias. At some locations of the islands, we indeed find that a rather broad peak showing up near zero-bias and the peak position may slightly deviate from zero-bias. (2) The proximity effect induces a 2D superconducting state which may resemble a state with the $p_x+ip_y$ symmetry. In this case, we should have an edge state constructed by the Majorana modes. A typical feature of this edge state is the existence of finite DOS at zero bias energy, but the momentum and spin of these edge states should have helical feature *(28)*. When the island has an appropriate size, its helical edge states from the perimeter just start to touch forming an interfered resonant state, we thus observe a ZBCP near the edge and also to the interior of the islands. (3) A magnetic moment, like interstitial iron, may exist just below this right island, which leads to the time reversal symmetry broken. This possibility might exist, but we think it is unlikely since all our Fe(Te,Se) samples have been annealed to remove the interstitial irons and achieve perfect superconducting transition before the scanning tunneling microscope (STM) measurements. In addition, it has no reason why these interstitial irons (if any) locate just below the islands showing the ZBCP, but not for others. Clearly, theoretical efforts are highly desired to unravel this puzzle.

Although the evidence of the ZBCP and probably the Majorana modes is robust in our experiment, we have still several key questions unanswered yet. Firstly, we are not sure what structural factors that control the appearance of the ZBCP. From our present experiment, we feel that the possibility of finding this effect on some round shaped islands with diameter of about 6-8 nanometers is high. However, what puzzles us is that we observed the ZBCP on the islands for the local structure of both distorted orthorhombic or hexagonal lattice. Secondly, it remains unclear what is the role played by the Fe(Te,Se) substrate. For the selection of the superconducting substrate, it is not clear whether the Fe(Te,Se) is unique or rather general. In previous experiments on Bi islands/film deposited on substrates of cuprate Bi2212 *(29)*, Nb *(30)* and 2H-NbSe$_2$ *(31,32)*, proximity effect induced superconductivity was observed, but without the trace of ZBCP. In any case, the evidence of the ZBCP in our present experiments is robust and the way for making this state seems quite simple. If this ZBCP is due to the Majorana modes, our results will open a new avenue for producing the topological superconducting state with Majorana modes, this will stimulate further theoretical and experimental efforts. For the simple way of making these states and the easy integrity in a solid state manner, our work may also help to braid these Majorana modes for the quantum computation.

**Acknowledgments:** We acknowledge the useful discussions with Qianghua Wang, Yoich Ando and Ziqiang Wang. We acknowledge Jin Si, Qing Li and Xiyu Zhu for the help during the growth of Fe(Te,Se). This work was supported by National Key R&D Program of China (grant no. 2016YFA0300400), National Natural Science Foundation of China (grant nos. 11534005, 11374144), and Strategic Priority Research Program of Chinese Academy of Sciences (grant no. XDB25000000). The sample Fe(Te,Se) was grown by M.C, W.D. The MBE growth and STM/STS measurements were finished by X.C, M.C, W.D, H.Y., and H.H.W. All authors discussed the results and contributed to the writing. H.Y and H.H.W take the responsibility for the final text. H.H.W has coordinated the whole work.

**Supplementary Materials:**

Materials and Methods

Figures S1-S3

References (*33*)

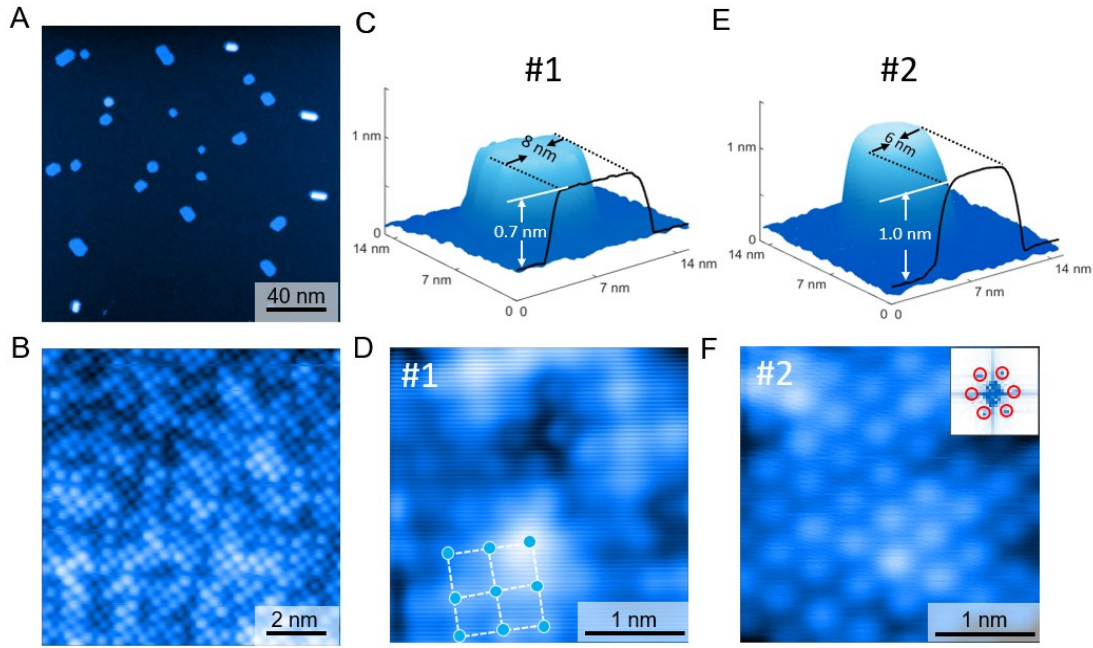

**Fig. 1 Topographic image of Bi islands on Fe(Te,Se) substrate.** (A) Topographic image of the Bi islands on the surface of Fe(Te,Se) single crystal. (B) Atomically resolved topographic image of Fe(Te,Se). (C and E) 3D plot of topographic images of two Bi islands with the height of 7 Å (C, Bi island #1) and 10 Å (E, Bi island #2). (D and F) Atomically-resolved topographic images measured on the Bi islands shown on C and E, respectively. The dashed frame in D illustrates the distorted orthorhombic structure. The inset in F shows the corresponding Fourier transformed topographic image of F, and one can see six Bragg peaks which correspond to the hexagonal structure with lattice constant of about 4.5 Å on the surface of Bi island #2. The topographic images were recorded at 1.8 K.

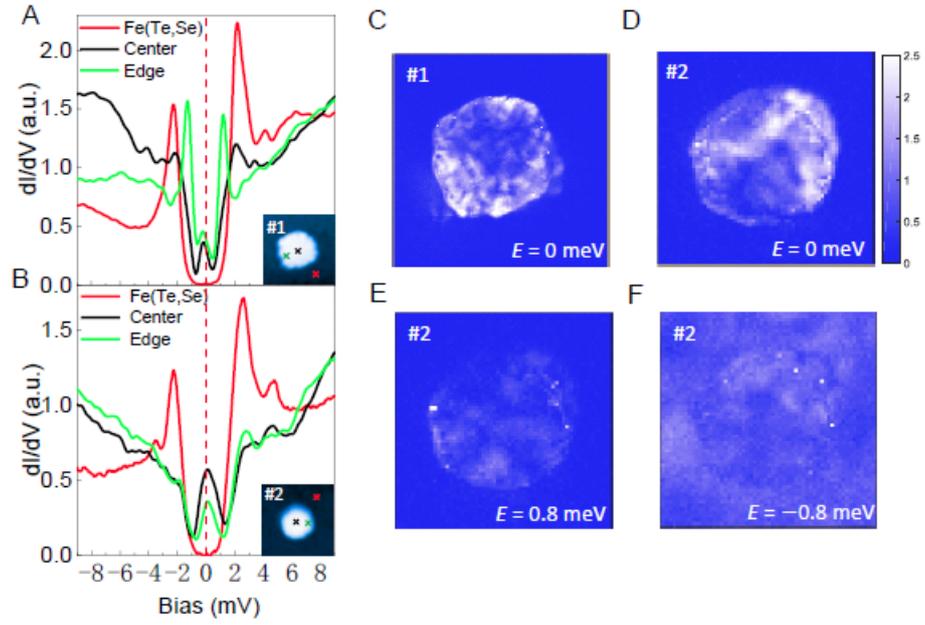

**Fig. 2 Tunneling spectra and differential conductance mappings.** (A and B) Tunneling spectra measured at the marked positions in the insets on Bi islands and nearby Fe(Te,Se) substrates at 0.4 K. A peak near zero bias can be observed within the superconducting gap energies on the spectra taken on the Bi islands. (C to F) Differential conductance mappings measured in the areas covering the Bi islands at various energies.

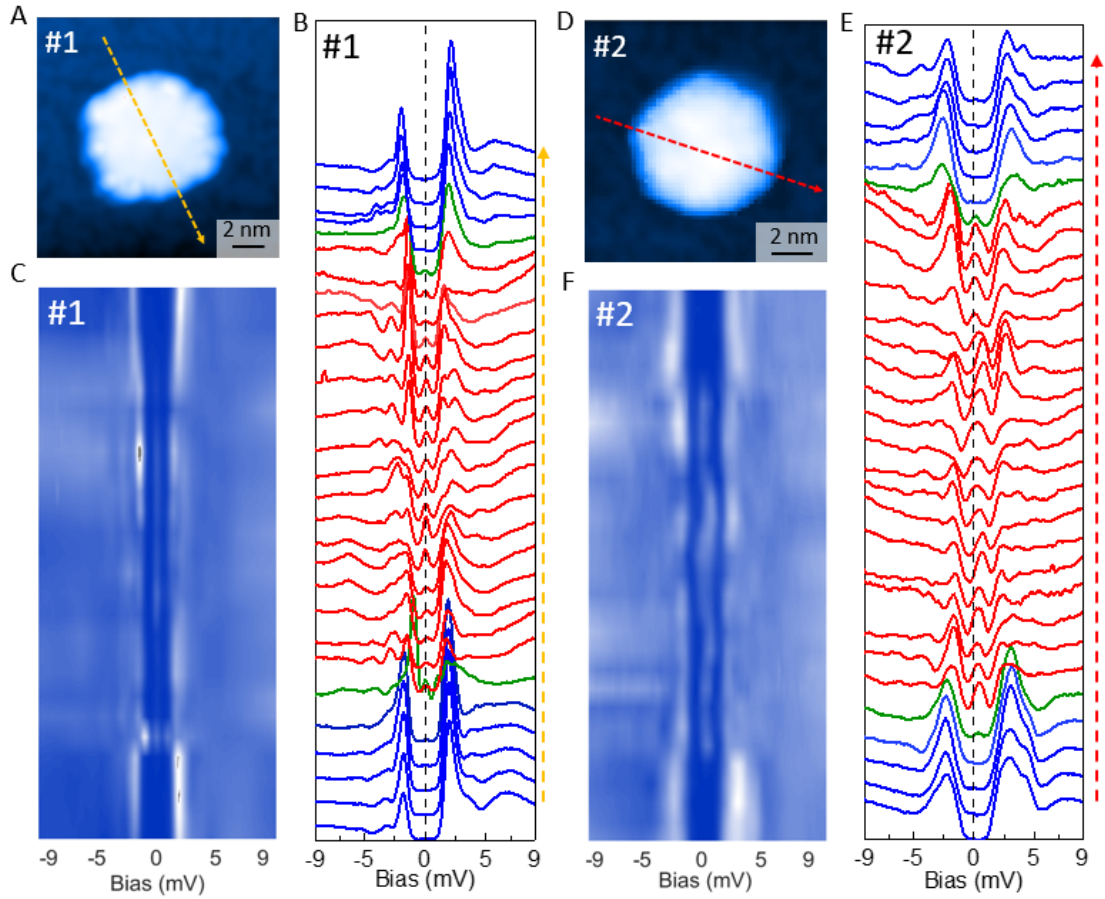

**Fig. 3 Spatially resolved tunneling spectra measured across Bi islands.** (A and D) Topographic images of the Bi islands. (B and E) Spatial profile of the spectra (with offsets) crossing the Bi islands along the arrowed dashed lines in A and D. The blue curves in B and E represent the spectra measured on Fe(Te,Se), and the red ones measured on Bi island, while the green ones measured near the edges. (C and F) Color plots of the tunneling spectra shown in B and E, respectively.

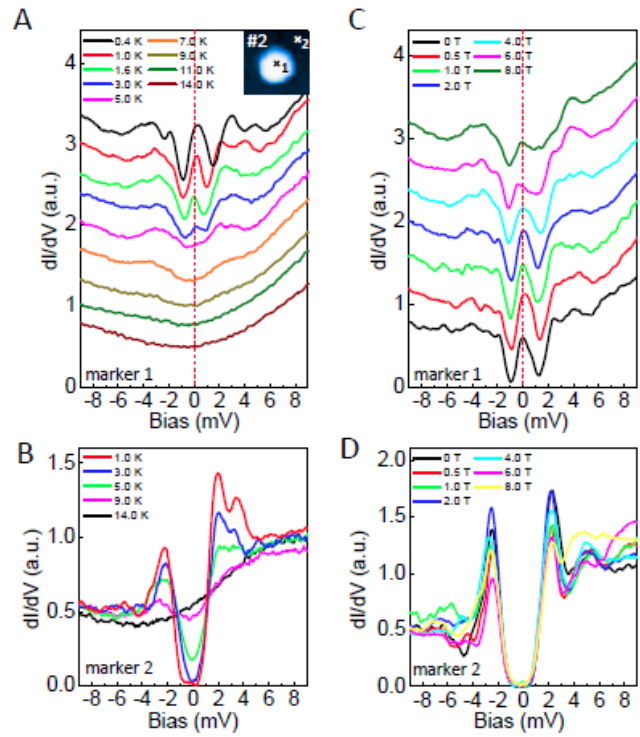

**Fig. 4 Temperature and magnetic field dependence of tunneling spectra.** (A and B) Temperature dependent tunneling spectra measured at marked positions on Bi island #2 and Fe(Te,Se). (C and D) Magnetic field dependent tunneling spectra.

# Supplementary Materials for

## Zero-Energy Modes on Superconducting Bismuth Islands Deposited on Fe(Te,Se)

Xiaoyu Chen*, Mingyang Chen*, Wen Duan, Huan Yang, and Hai-Hu Wen†

Correspondence to: hhwen@nju.edu.cn (H. H. W.).

**Materials and Methods**

The single crystals of Fe(Te,Se) are used as substrates for the growth of Bi islands. The Fe(Te,Se) samples are grown by self-flux method *(33)*. In order to reduce the interstitial Fe atoms, the Fe(Te,Se) samples are annealed in $O_2$ atmosphere at 400 °C for 20 hours and then quenched in liquid nitrogen. Before the deposition of Bi atoms, the Fe(Te,Se) samples are cleaved in the chamber with ultrahigh vacuum about $1\times10^{-10}$ Torr. The Bi atoms (99.999%) are deposited on the cleaved surface of Fe(Te,Se) at room temperature by molecular beam epitaxy technique by the CreaTec effusion cells.

The STM/STS measurements in this work are carried out in a scanning tunneling microscope (USM-1300, Unisoku Co. Ltd.) with ultrahigh vacuum, low temperature and high magnetic field. The tunneling spectra are measured by using a lock-in technique with amplitude of 0.3 meV and frequency of 931.773 Hz.

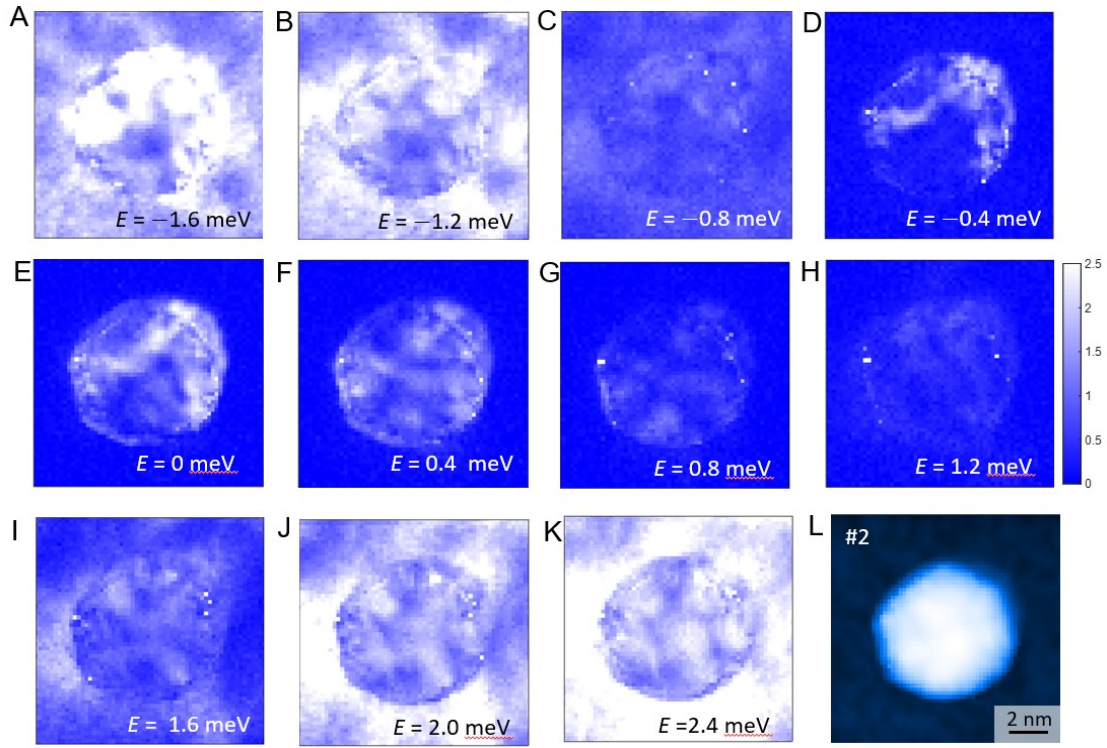

**Fig. S1. Differential conductance mappings on island #2.** (A to K) Differential conductance mappings at various energies ($I_t$=50 pA, $V_t$=10 mV). (L) Corresponding topography of the view for measurements.

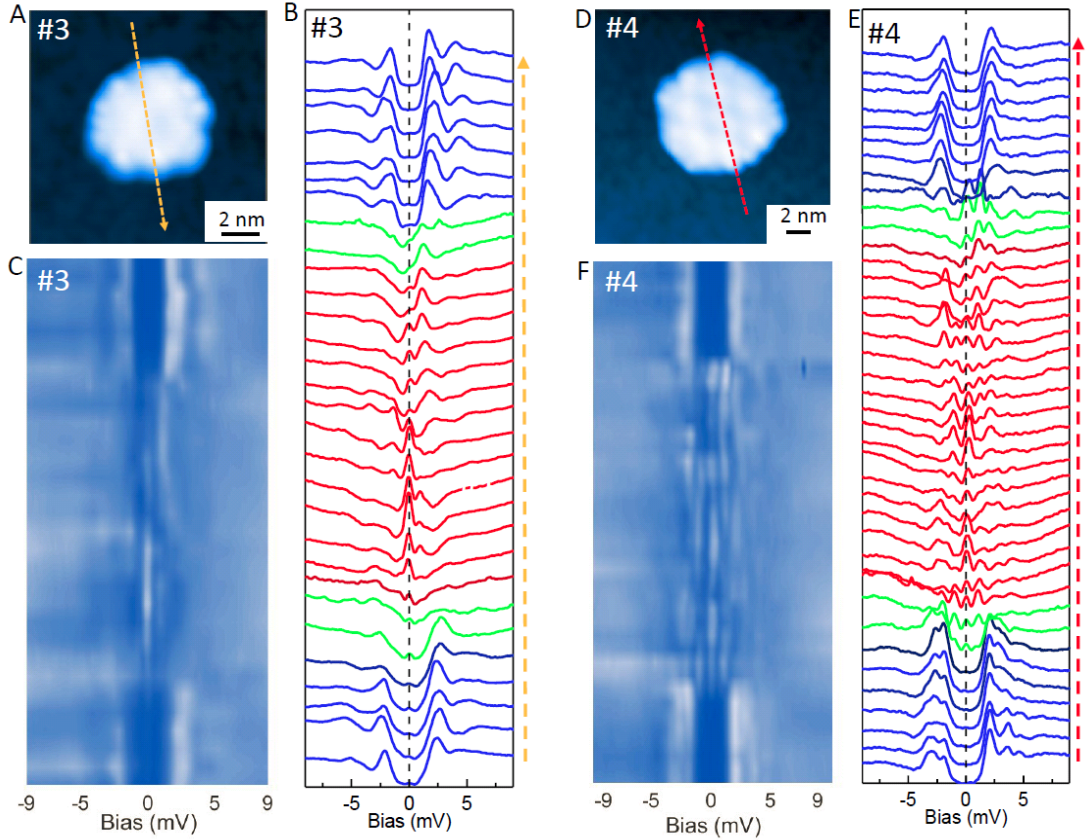

**Fig. S2. Spatial evolution of tunneling spectra measured across other Bi islands.** (A and D) Topographic images of the Bi island #3 and #4. (B and E) Spatial profile of the spectra (with offsets) crossing the Bi islands along the arrowed dashed lines in A and D. The blue curves in B and E represent the spectra measured on Fe(Te,Se), and the red ones measured on Bi island, while the green ones measured near the edges. (C and F) Color plots of the tunneling spectra shown in B and E, respectively.

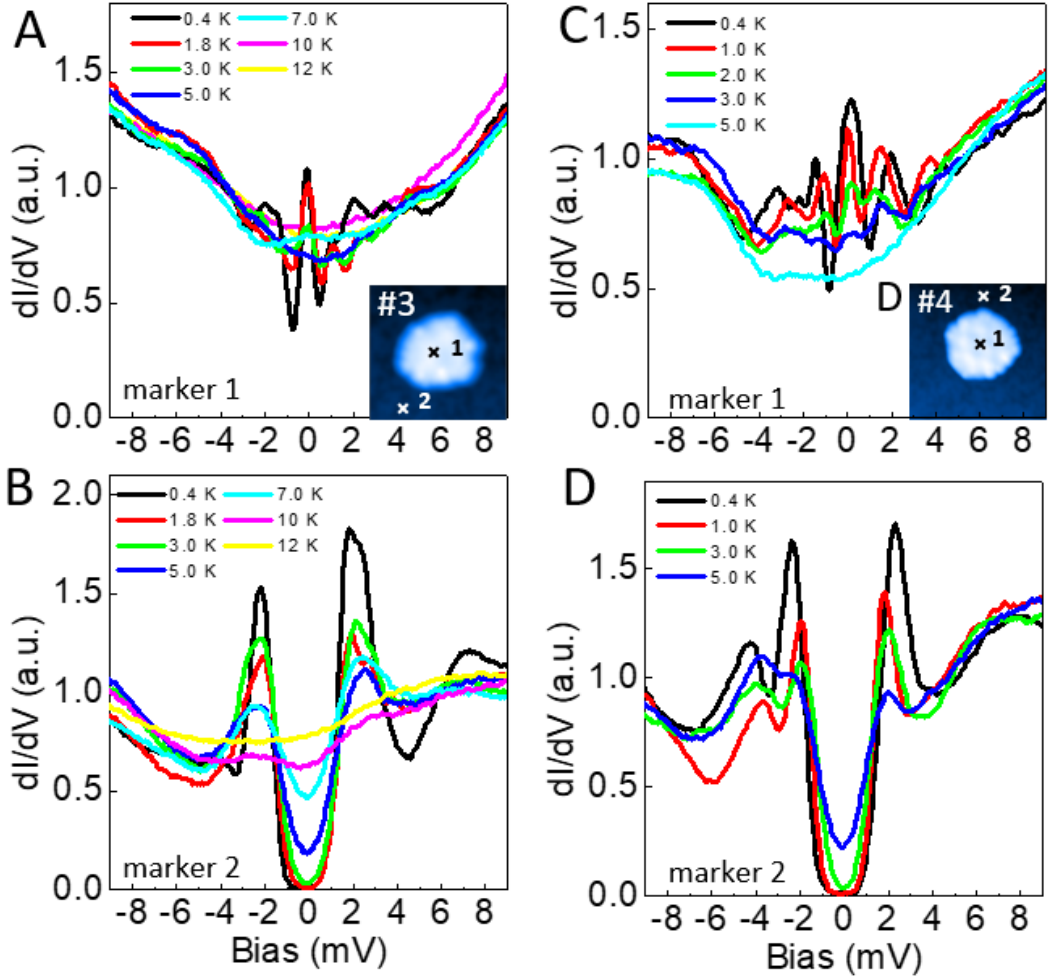

**Fig. S3. Temperature dependent tunneling spectra. (A to D)** Temperature dependent tunneling spectra measured at marked positions on Bi island #3, #4 and Fe(Te,Se).